\documentclass[preprint,aps,showpacs,showkeys,amsmath,amssymb]{revtex4}

\usepackage{graphicx}
\usepackage{latexsym}
\usepackage{dcolumn}
\usepackage{bm}

\begin{document}

\preprint{Massey/CSTN-036}

\title{
Ising Model Scaling Behaviour on z-Preserving Small-World Networks}

\author{K.A.Hawick} \email{k.a.hawick@massey.ac.nz}

\author{H.A.James}  \email{h.a.james@massey.ac.nz}

\affiliation{
Institute of Information and Mathematical Sciences\\
Massey University, North Shore 102-904, Auckland, New Zealand
}

\date{\today}

\begin{abstract}
  We have investigated the anomalous scaling behaviour of the Ising
  model on small-world networks based on 2- and 3-dimensional lattices
  using Monte Carlo simulations.  Our main result is that even at low
  $p$, the shift in the critical temperature $\Delta T_c$ scales as
  $p^{s}$, with $s \approx 0.50$ for 2-D systems, $s \approx 0.698$
  for 3-D and $s \approx 0.75$ for 4-D.  We have also verified that a
  $z$-preserving rewiring algorithm still exhibits small-world effects
  and yet is more directly comparable with the conventional Ising
  model; the small-world effect is due to enhanced long-range
  correlations and not the change in effective dimension. We find the
  critical exponents $\beta$ and $\nu$ exhibit a monotonic change
  between an Ising-like transition and mean-field behaviour in 2- and
  3-dimensional systems.
\end{abstract}

\pacs{89.75.Hc, 89.75.Da, 64.60.Fr, 64.60.Ht }

\keywords{Ising; scaling; small-world}

\maketitle

\section{Introduction}
\label{sec:introduction}

Small-world networks~\cite{Watts+Strogatz} are now an important class of complex network and have been used for many
purposes ranging from social network models~\cite{Klemm-SocialNetworks} to quantum gravity~\cite{WanOnSmallWorldLQG}.
Complex networks have many unusual properties~\cite{NewmanComplexNetworks} but one of the most valuable aspects of the
small-world model is that it allows a parameterized interpolation between those properties exhibited by a random
graph~\cite{Bollobas} and those by regular lattices.  Understanding the process that lead to small-world networks is
non-trivial~\cite{Mathias+Gopal,NewmanSmallWorldModels} and one important approach involves comparing the critical phase
transitional behaviour of small-world systems with well-studied phase transition models.  The Ising model provides a
valuable test model for studying phase transitions on various networks.  Study of the 1-D Ising
model~\cite{LopesExactIsing} has led to some insights and relatively recent work by Herrero~\cite{Herrero,Herrero2} and
others~\cite{HastingsSmallWorld} raises some interesting questions about the behaviour of the Ising model on small-world
rewired lattices.

The Ising model can be adapted in a number of ways such as adding long-range weak
interactions~\cite{Zhang+NovotnyOnLongRange}, however we are interested in preserving the underpinning model structure
and consider solely the effect on spatial distortions and in particular shortcuts.  It is still unclear whether real
physical materials exhibit small-world magnetic properties~\cite{NovotnyEtAlOnNanomaterials} but one can speculate about
small-world ``shortcut'' effects that might arise when effectively one-dimensional structures such as protein chains are
folded.  Two-dimensional crumpled sheet systems embedded in a three-dimensional space could also exhibit real-space
shortcuts.  We might further speculate that higher dimensional systems such as quantum gravity models, that are not
restricted to three Euclidean dimensions, could also exhibit physically relevant small-world transitional behaviour.

The Ising model is usually formulated on a regular d-dimensional hyper-cubic lattice where each magnetic spin variable
is connected to $2 \times d$ nearest neighbouring sites.  The  model Hamiltonian is  usually written in the form:
\begin{equation}
H = - \sum_{i \neq j} J_{ij} S_i S_j
\end{equation}
where $S_i = \pm1$, $i=1,2,...N$ sites, and $J_{ij}$ is $|J| = 1 / k_BT$ is the ferromagnetic coupling over neighbouring
sites $i$,$j$ on the network.

The model has a critical temperature of $T_c = \frac{1}{J_c k_B}= 0$ in one dimension, but displays finite transition
temperatures due to the spontaneous magnetization effects in higher dimensions.  Specifically in two- and
three-dimensions the Ising-type phase transitions have been very well studied and the critical temperatures are known
exactly in two dimensions~\cite{Onsager} and approximately from computer simulations such as~\cite{MCRG} in three
dimensions.  In systems of four dimensions and higher a finite temperature phase transition still occurs, but the nature
of the transition is well described by mean-field theory.  Values for the transition temperature in these higher
dimensional systems are also known.

The regular lattice Ising model can be perturbed in a number of controlled ways: by removing bonds or ``damaging'' the
system; by rewiring the bonds; or by adding extra bonds.  Herrero and others have employed the Watts-Strogatz rewiring
model which preserves the number of total bonds, but does not preserve the number of bonds connecting a given site.
Svenson and Johnson explored damage spreading in small-world Ising systems~\cite{Svenson+Johnson}, again using the
Watt-Strogatz approach.

Generally, the Ising model phase transitional temperature is systematically shifted by these damaged, rewired or added
links.  In the case of the small-world rewiring, individual sites can become connected to sites very far away
physically.  These long-range bonds encourage and help the long-range correlations that manifest the spontaneous
magnetization and hence give rise to the peculiarly Ising-like critical phenomena.  The Ising model critical temperature
rises as long-range order is encouraged by the rewiring.  Physically, it is easier for the system to maintain long-range
order against the thermal disordering effects of higher temperatures than it would be without the rewired long-range
bonds.  This can be measured as a monotonic dependence of the critical temperature $T_c(p)$ on the small-wiring rewiring
probability $p$.

It has been speculated that under the Watts-Strogatz small-world rewiring this shift in $T_c$ is partially due to the
effective change in the dimensionality of the system as individual sites can have more or less than $ z = 2 \times d$
bond-connected neighbours.  In this paper we explore an alternative small-world rewiring model that adjusts pairs of
bonds and is able to preserve exactly the coordination number $z$ for all spin sites.

This means that as far as individual spin-sites are concerned their environment is locally identical to the conventional
$z=2 \times d$ lattice-based Ising model.  We explore the behaviour of the Ising model under such a rewiring and compare
it with the Watts-Strogatz rewiring studied by Herrero and others.  We are able to show that the small-world effect is
not dependent on an effective dimension change but solely on the long-range nature of the interactions themselves.

In our model, spins still have precisely $z$ ``nearest neighbour'' sites to which they are directly connected and with
which they are topologically directly coupled via the Ising model coupling parameter $J$.  However in terms of a
physical space interpretation, the rewired links mean that nearest neighbouring sites can now be at arbitrary physical
distances apart.  This can be interpreted as filling physical space with ``worm-holes'' or applying an elaborate
manifold folding of physical space.

Like that of~\cite{Herrero} our study is based on Monte Carlo simulations of the Ising system on rewired lattices.  We have
been able to study larger systems than prior work for both two- and three-dimensions and have also made some preliminary
studies of four- and five dimensional systems.  Of particular importance is the need to study systems that are large
enough to support the so called small-$p$ regime.  It transpires that the power law dependence on important properties
of the model on the rewiring probability parameter $p$ requires an analysis across logarithmic system sizes and length
scales.  Consequently we have needed to simulate large Ising systems of up to $1024^2$ and $384^3$.

Equilibration requirements for large system sizes over a large parameter space set of $p$ and $T$ values needed a faster
Monte Carlo simulation algorithm than the local Metropolis method used by prior small-world Ising work.  We therefore
have explored non-local cluster-based updating methods such as that of Wolff~\cite{Wolff} for our simulations.  We
certainly expected the Wolff cluster method to perform entirely correctly on a $z$-preserving rewiring model, and
indeed it does.  We have also, however, verified that it produces results on the non-$z$-preserving Watts-Strogatz rewired
system that are consistent with those using the localised Metropolis updating method.

The main question of interest is how the critical phenomenological behaviour~\cite{CriticalPhenomena} of the model
changes due to the small-world links.  Approximate theoretical models for the $p$-concentration dependence of $T_c$, such
as those given in~\cite{Meilikhov+Farzetdinova}, depend on arguments based on domain wall formation energies in the
Ising model.  In the work reported here, we show that the critical temperature is monotonically shifted upwards from its
regular lattice value in $d = 2,3,4,5$ and that it is very-well characterised by a power law in $p$.  We also consider
the critical exponents $\beta$ and $\nu$ and our data shows a gradual but monotonic change with increasing $p$ from an
Ising-like phase transition to a mean-field-like transition in both two- and three-dimensions.

Although a great deal of practical lore exists in the literature concerning the operational methods of obtaining
critical values for the regular Ising model, it is more difficult to obtain corresponding values for small-world systems
due to less well studied $p$-regimes and the need for appropriate averaging of rewired lattice configurations.  We
therefore describe our computational method in detail in section~\ref{sec:method}.  Our main results concern the nature
of the rewiring models and the interpretation of the parameter $p$ and in section~\ref{sec:rewiring} we describe our
rewiring algorithm.  In section~\ref{sec:results} we present some results of the Monte Carlo simulations and for some
static network properties such as the maximum path length in a rewired lattice.  Finally we suggest some conclusions and
areas for further simulation in section~\ref{sec:conclusions}.

\section{Computational Method}
\label{sec:method}

Herrero noted the uncertainty as to whether his simulated system sizes were sufficiently large to truly be able to
explore the ``small-$p$'' regime.  We have experimented with a variety of system sizes in both two- and three-dimensions.  We
believe a useful rule of thumb is that at least 100 of the system's bonds need to be rewired to have a reliable and
measurable effect.  This places a practical lower limit on the p-values that can be explored, given the practical upper
limits on system sizes that can be reliably simulated.

In view of the need to simulate large systems sizes, for long measurement periods and for many different $p$-values near
criticality, we felt it was important to investigate fast algorithms such as cluster updating methods.  The Wolff
cluster updating algorithm~\cite{Wolff} carries the Ising spin configuration to a new point in phase space by
constructing a cluster of {\em like} spins with an appropriate probability, based on the coupling (and hence the simulated
temperature).  This algorithm is particularly effective at temperatures near the critical point as it can flip very
large spin clusters and effectively overcomes the critical slowing down of a local update method such as Metropolis.  In
this work we are interested in equilibrium properties, not dynamical ones and therefore the time evolution properties of
the update algorithm need not resemble any physical process.  We verified our implementation of the Wolff algorithm
against implementations of both Metropolis and Glauber~\cite{Glauber} Monte Carlo updates.  For the work reported here,
we note no discernible difference in quality of equilibrium properties estimated.

\subsection*{Computational Resourcing Issues}

Although the work was carried out on a mix of 32-bit and 64-bit microprocessors, we found that it is still only
tractable to simulate systems that comfortably fit within approximately 2 GBytes of memory.  This is the addressable limit
attainable using a 32-bit signed integer, and although a bit-based model such as the Ising model can be implemented
using a compact storage scheme, for implementations of the Wolff cluster algorithm we require a full integer for each of
the $N$ spin sites.  Furthermore, when investigating non-regular networks we need to explicitly store the neighbour
addresses for each site.  For our code, we optimise for speed using a storage scheme that records the neighbour arcs for
every site -- thus storing a bond's source and destination site twice.

In the case of a rewiring model that allows different sites to have different coordination numbers, this scheme gives
rise to a storage budget of one spin-bit; one mark-bit; one coordination number (integer) and on average $2\times d$
site addresses per spin site.  The practical upshot of this is that we can, in principal, simulate Ising systems of up
$384^3$ sites.  In practice however, even with the Wolff algorithm to assist with equilibration and decorrelation near
the critical temperature, we found that the processors available to us for this work would take over 1 week of wall-clock
time to attain a useful measurement for one $T-p$ pair.  Consequently we found that size limits of around $N = 256^3$
(around 16 million sites) were more practical.  We anticipate that larger sizes (approaching $512^3$) will be practical
with 64-bit address spaces and the next generation of processor speeds.

In applying the Binder cumulant method we found it useful to run triples of systems size, such as N=$224^3$, $240^3$,
$256^3$ in three dimensions, or $992^2$, $1024^2$ and $1056^2$ in two dimensions.  Although only two cumulant curves are
needed to obtain an intersection, three combine to also give a measure of uncertainty.

\subsection*{Monte Carlo Algorithmic Issues}

The Wolff cluster algorithm is not efficient when performing an initial quench from a hot or random spin configuration,
since the clusters it builds tend to be very small.  Given the unknown dependencies on the parameter $p$ we erred on the
side of caution and generated a completely independent start configuration for each experimental $T-p$ pair
investigated.  Typically we quenched a hot starting spin configuration to finite $T$ using on average 1000 Metropolis
hits per site.  Although the debate in the literature over the effect of sweeping artifacts seems now resolved, we
avoided any bias accidentally introduced by sweeps by performing the Metropolis equilibration site hits in a random
order.  So in effect we apply $N \times 1000$ randomly chosen local Metropolis hits following the quench.  Since we are
typically interested in behaviour near $T_{c}^{p}$, we typically used the Wolff cluster method predominantly during
measurement, but again to err on the side of caution, we used a hybrid update step consisting of 100 Wolff cluster hits,
followed by $N$ local Metropolis hits for each measurement.

In determining the Binder cumulant and other statistics, we typically made 12 million measurement steps, dividing them
into blocks of 1 million, and discarding the first two blocks as additional equilibration.  For the smaller system
sizes, we found that 100-150 thousand measurement steps were sufficient.  The data blocking approach gave us a measure of
uncertainty in each cumulant value.

For the large systems sizes used it appears that good self-averaging properties hold and the results are largely
independent of the random rewiring pattern chosen for a particular run at a particular $p$ value.  For small-$p$ values
and also for smaller system sizes we found a perceptible experimental spread of results and it was necessary to repeat
runs at the same $T-p$ value.  In the work reported here we typically combined measured cumulant values from 16 or 25
completely independent runs.  Study of the statistics indicated a satisfactory central-limiting behaviour and a ready
calculation of uncertainties in the cumulants so obtained.

In most of the work reported here, we used the lagged-Fibonacci random number generator of Marsaglia~\cite{Marsaglia}.
For some simulations on 64-bit platforms we used an implementation of the Mersenne-Twistor generator
algorithm~\cite{MersenneTwistor}.  We are not aware of any concerns regarding periodicity or correlations due to either
of these generators.

\section{rewiring Algorithms}
\label{sec:rewiring}

The value of a small-world rewiring procedure is that the amount of space folding can be specified statistically by a
single parameter $p$, with the extreme value $p=0$ corresponding to a normal periodic hyper-cubic geometry lattice.  The
case $p=1$ corresponds to a random network of spins.

Herrero and other authors report using a network rewiring algorithm that maintains an average
coordination number $z$ for each spin site, but that does allow individual spins to have a lesser or
greater $z$ value.  This small-world rewiring algorithm is essentially that described by Watts and
Strogatz and works by randomly selecting $p \times N \times d$ of the $N \times d$ original regular-lattice bonds
and reassigning them to link random spin sites.

This algorithm can be implemented as:
\begin{enumerate}
\item choose a spin site {\bf A} at random
\item choose one of its $2 \times d$ existing neighbour {\bf B} at random
\item choose another (distinct) spin site {\bf C}
\item re-wire A to C, disconnecting A from B
\end{enumerate}

In this rewiring model $p$ is the probability that any of the $z \times N$ regular bonds has been rewired.  It can be
verified experimentally that a requested $p$ value from the rewiring algorithm has been implemented by performing a
bond-by-bond comparison with each bond's regular lattice endpoints.
  
One difficulty with this model is that the system is also influenced by percolation transition effects.  Even at
small-$p$ values individual spin sites have a finite probability of becoming completely disconnected from the rest of
the system.  In practice the system will have a finite number of monomer and dimer spin sites, separate from the one
giant component.  At large $p$ values (greater than $0.1$ for example) this is a serious effect and the spin system is
typically fragmented into more than one major component.  Although this can be compensated for by only making
measurements on the largest component, it is an operational annoyance as well as seriously worsening finite-size effects
that impinge on the calculation of the critical temperature and exponents.  Above the bond percolation threshold of $p =
0.5$ the system is nearly always very fragmented and consists of a number of similar-sized components.

A major aim of our work was to investigate the effect of a $z$-preserving network rewiring algorithm both in terms of
how it compared to the Watts-Strogatz edge-rewiring algorithm and also how to interpret its probabilistic rewiring
parameter $p$.  To preserve coordination number $z$ exactly and for all spin sites it is necessary to re-wire the
regular lattice in terms of pairs of edges.

Our network is constructed from the starting lattice using a similar rewiring procedure.  We modify the bonds of each site so
that it still has degree $z = 2 \times d$ bonds per site in dimension $d$, but that they may link, with probability $p$,
to a randomly chosen site elsewhere in the original lattice.  We ensure each site links to other sites at most once, and
there are no self-bonds in our system.  This is feasible below the percolation threshold.  Specifically, our procedure
is:

\begin{enumerate}
\item choose a spin site {\bf A} at random
\item choose one of its $2 \times d$ existing neighbours {\bf B} at random
\item choose another (distinct) spin site {\bf C}
\item choose one  of its neighbours {\bf D} at random 
\item ensure {\bf A}, {\bf B}, {\bf C} and {\bf D} are all distinct to avoid  self  bonds and multiple bonds
\item re-wire  {\bf A} to {\bf C}  and {\bf B} to {\bf D}, thus exactly preserving $z$ for all of {\bf A}, {\bf B}, {\bf C}
  and {\bf D}
\end{enumerate}

Repeating this procedure $p \times N \times d$ times achieves a rewiring of the regular lattice to have an effective
rewiring probability $p^{*}$ which can be subsequently measured by comparing spin-site neighbour-lists to that of the
regular lattice.

This algorithm has the additional desirable property that it is guaranteed not to fragment the lattice into multiple
components since each site still connects to $z$ distinct other sites.

\begin{figure}[htbp]
\begin{tabular}{cp{0.4cm}c}
\includegraphics[angle=0,width=4cm]{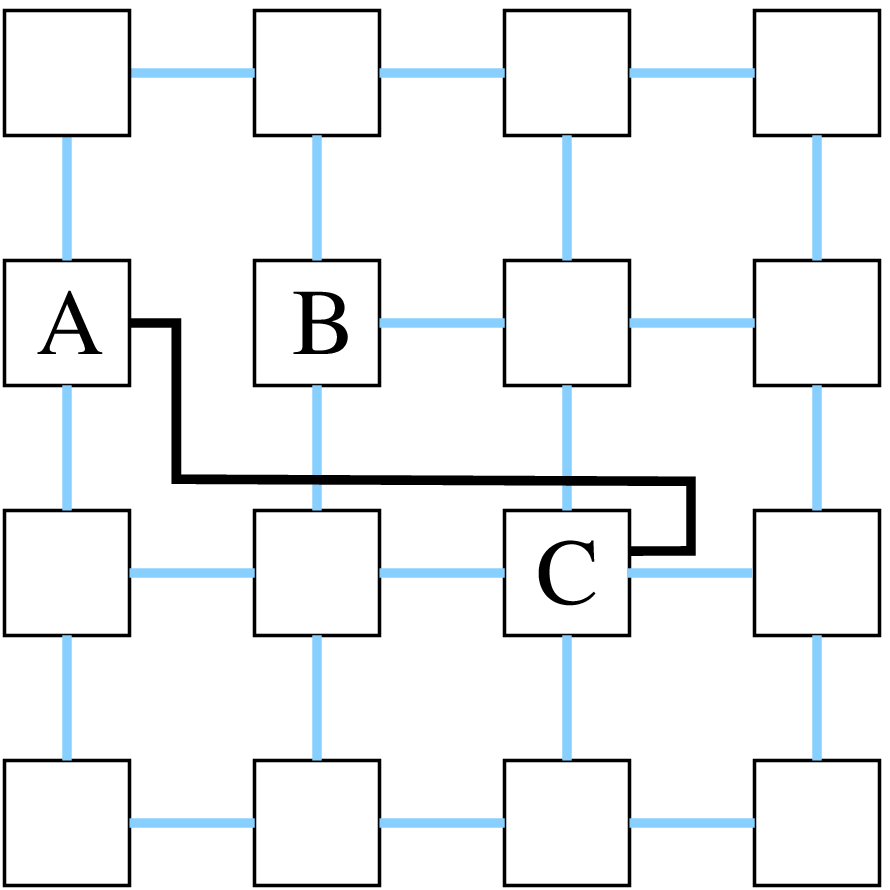} & &
\includegraphics[angle=0,width=4cm]{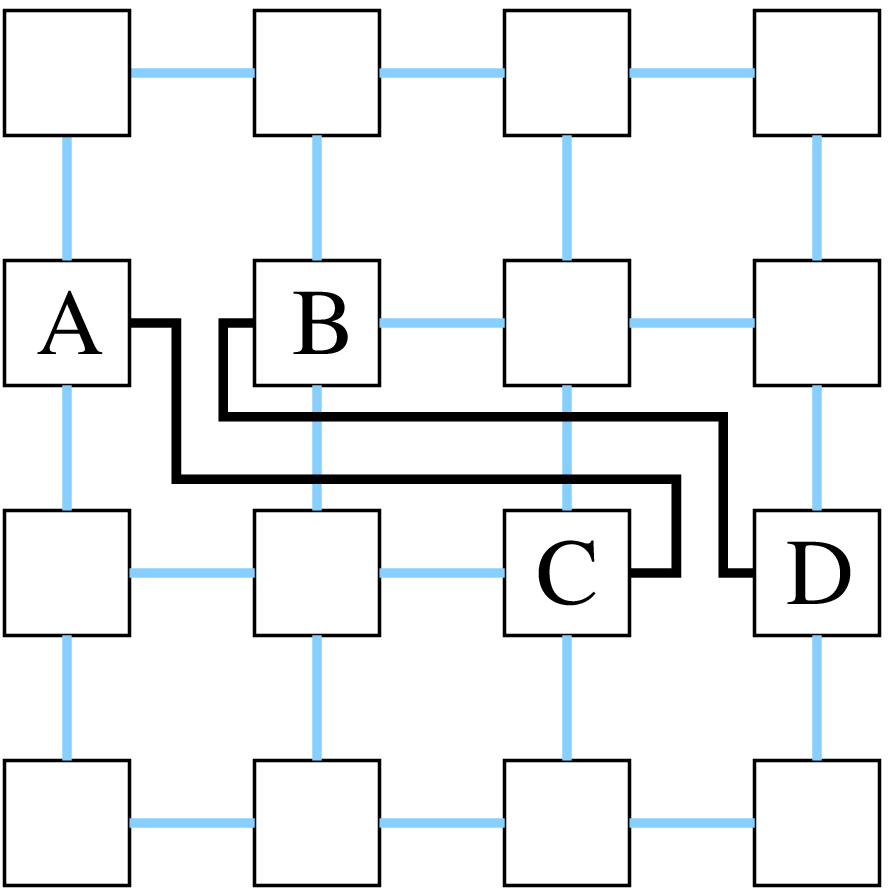} \\
a) & & b) \\
\end{tabular}
\caption{\label{fig:rewiring} Different rewiring models: a) Watts-Strogatz non-$z$
  preserving model, b) Our $z$-preserving rewiring model. }
\end{figure}

Figure~\ref{fig:rewiring} shows an example of the two rewiring models for a regular square lattice substrate.  

In practice, the achieved or effective rewiring probability $p^{*} < p$ by an amount that is of order $p^2$ since our
algorithm allows a site to be ``re-rewired''. There is therefore an exclusion effect that gives rise to a correction
term in $p$.  Since, $p^{*}$ is readily measured, this is only an operational inconvenience and a chosen effective
$p^{*}$ can be set with an appropriate choice of $p$ given the solvable quadratic relationship between them and
experimentally fitted coefficients.

Over and above this exclusion correction however, it is important to note that {\em pairs} of bonds have been rewired.
This pairing leads to a factor of two in the effective value of $p$ for our model when we compare it with the
Watts-Strogatz rewiring model.  This is explained when we consider the actual effect of a pair of rewired links
connecting sites {\bf A}, {\bf B}, {\bf C}, {\bf D} as shown in figure~\ref{fig:rewiring}.  If {\bf A} and {\bf B} were
originally neighbours, and so were {\bf C} and {\bf D}, and now {\bf A}, {\bf B} are connected via a long-range pair of
links to {\bf C}, {\bf D} then the net effect is really only as if there were one long range link.  This manifests itself
in the role of $p$ for our $z$-preserving rewiring model.  In comparing the two models, we need to measure the actual
$p$ implemented by comparing with the regular lattice bond end-points.  Then we should treat the effective $p$ for the
$z$-preserving model as half the effective $p$ achieved by the Watts-Strogatz model.  This is further evidence that it
is the long-range nature of the rewired bonds that is behind the small-world critical shifts in the Ising model -- the
local topological details (such as which particular neighbour of {\bf A}, {\bf B}, etc. was rewired) are much less
important.

\section{Results and Discussion}
\label{sec:results}

The effect of the small-world network rewiring is to add long-range interactions across the lattice.  This enables the
formation of correlations across the spin sites and consequently makes it easier for the Ising system to maintain
spontaneous magnetization at a higher critical temperature than it otherwise would.  This effect is not dissimilar to
the finite-size effects of simulating {\em periodic} lattice models.  The periodicity means that the system is able to
support spin correlations at a higher temperature that it otherwise could and consequently the critical temperature is
shifted to a higher value than that of a real system in the thermodynamic limit. A key step to understanding the
small-world effect is to quantify the resulting shift in the critical temperature for the different network rewiring
models.

\subsection*{Binder Cumulant Analysis}

We compute the temperature shift $\Delta T_c = T_{c}^{p} - T_{c}^{p=0}$ based on the critical temperature measured for a
small-world rewired Ising model system, compared with that of an unperturbed system on a regular lattice.

\begin{figure}[hbt]
\begin{center}
\includegraphics[angle=-90,width=9cm]{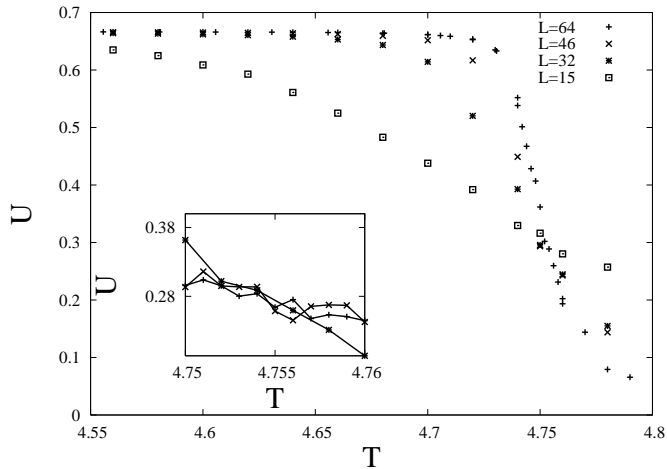}
\caption{\label{fig:binder} Binder Cumulants }
\end{center}
\end{figure}

The  critical temperature of a Monte Carlo simulated system can be calculated using the fourth-order Binder
cumulant method~\cite{Binder+LandauCumulant,Binder+Heermann}.  The  cumulant:
\begin{equation}
U_N = 1 - \frac{\langle M^4 \rangle_N}{3 \langle M^2 \rangle^2_N}
\label{eq:cumulant}
\end{equation}
is defined at different network sizes $N$, where $N=L^d$ based on the edge length $L$ of our substrate lattice.  The
cumulant shows a transition edge at the critical temperature.  The cumulant curves for different $N$ (or $L$) values
coincide at the critical temperature and this gives a way of extrapolating to the thermodynamic limit from relatively
small sized simulations.  Figure~\ref{fig:binder} shows the form of the cumulant and how it has a sharper transition
edge at larger system sizes.  A practical method of obtaining a critical temperature is to simulate at least three
different system sizes and by fitting straight lines to the linear region of the cumulant curve around the critical
temperature, calculate the intercept and an uncertainty estimate.  For all the work we report in this paper we used at
least three system sizes.  For small systems multiple independent rewiring configurations can be sampled.  For very
large systems needed for estimating low-$p$ behaviour it is impractical to run more than a few independent samples and
consequently the measurement uncertainties are very much greater.  Generally we have been able to estimate the critical
temperatures for two- and three-dimensional systems to around four significant figures, to three significant figures for
four-dimensional systems.  Our five-dimensional system simulations are limited to qualitatively showing that a
small-world effect takes place.

\begin{figure}[hbt]
\begin{center}
\includegraphics[angle=-90,width=9cm]{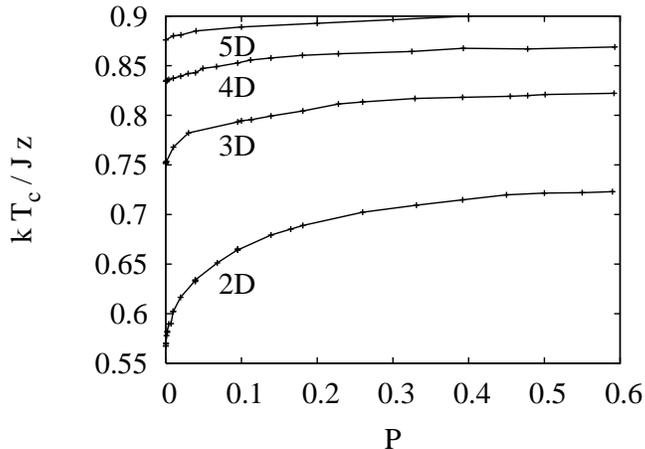}
\caption{\label{fig:kTperJz} Variation of the Critical Temperature against small-world (bond pair) rewiring probability for
  Ising model in 2, 3, 4 and 5 dimensions, scaled by the underpinning regular lattice coordination number
  $z$.  Error-bars are comparable to symbol sizes and lines joining points are guides to the eye only.
}

\end{center}
\end{figure}

Generally, obtaining the shifted critical temperatures for different $p$-values was an iterative procedure.  We employed
small lattice sizes to home in on approximate locations then used progressively larger system sizes to refine precision
and uncertainties.  Figure~\ref{fig:kTperJz} shows the qualitative behaviour of the critical temperature as it varies
with $p$ for different dimensionalities.

\begin{figure}[hbt]
\begin{center}
\includegraphics[angle=-90,width=9cm]{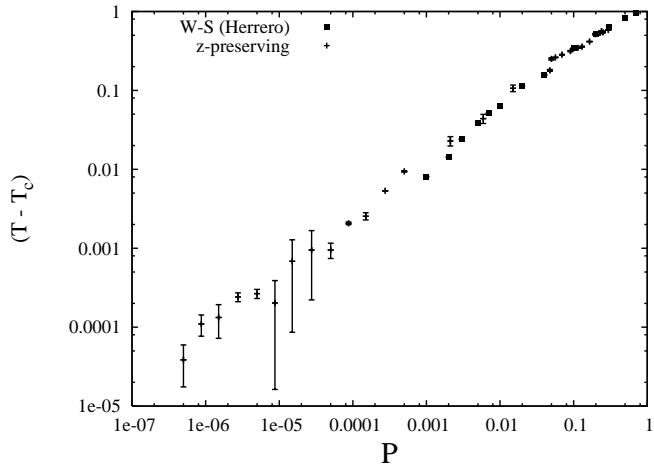}
\caption{\label{fig:collapse} Collapse of  Wiring model 1 data onto Model 3 (actual p) after factor of two scaling }
\end{center}
\end{figure}

Analysis of the shift $\Delta T_c = T_{c}^{p} - T_{c}^{p=0}$ in the critical temperature from the regular lattice value
of shows a power law of the form $\Delta T_{c}^{(p)} \approx p^s$.  This is shown on a logarithmic scale for three
dimensions in figure~\ref{fig:collapse} where we have included all our three-dimensional data for both the
Watts-Strogatz rewiring model and our $z$-preserving model (after suitable corrections to the meaning of $p$), as well as
data taken from~\cite{Herrero}.  We believe that within the bounds of experimental uncertainty these all agree.

This confirms that well below the percolation limit, where the system is essentially one single component, the behaviour
is independent of the rewiring model.  We have been able to extend the simulation results of Herrero to much larger
system sizes and hence much smaller values of $p$.

We studied both the Watts-Strogatz and our $z$-preserving rewiring models in detail on small-world systems constructed
from 2-D, 3-D and 4-D hyper-cubic lattices.  For each dimension we studied at least three decades of $p$ values and in
the case of 3-D systems we studied six decades of $p$ down to $10^{-6}$ on systems of up to $N \approx 384^3$.  A
preliminary study of a 5-D system also indicates that the small-world shift in $T_c$ does take place, although our 5-D
data is as yet insufficient to determine a useful value of $s$.

For $p \lesssim 0.1$ we are below the percolation regime and the system consists of one large component for the W-S
model.  In this regime our rewiring model and the W-S rewiring model are in agreement within the limits of experimental
error, and furthermore agree closely with the 3-D data reported for a W-S rewiring model~\cite{Herrero}.  Combining all
data we find $s \approx 0.698 \pm 0.002 $.  Although we have less data for 2-D systems, we investigated larger system
sizes (up to $1024^2$) than Herrero, and do not find the tail-off he reports for small-$p$.  We find a very good fit for
$s \approx 0.50 \pm 0.001$ for 2-D systems.

We conclude that the $\Delta T_{c}^{(p)} \approx p^s$ power-law is a very good description of all the data and any
deviation from it is indeed due to finite-size limitations that do not properly achieve the small-$p$ regime as Herrero
correctly suggests for his data.

\subsection*{Critical Exponent $\beta$}

It remains to explore the critical exponents of the simulated small-world system to determine in what manner the
system transitions between the Ising-type transitions at $p=0$ in two- and three-dimensions, and the mean-field like
transitions at high-$p$ values.

The critical exponent $\beta$ describes how the Ising model order parameter -- its magnetization $M$ -- diverges close to
the critical temperature.  It is usually defined by:
\begin{equation}
\langle M \rangle \sim |T_{c}-T|^{\beta}, T < T_{c}
\end{equation}
and we define the reduced temperature $t \equiv T_{c} - T$ so that $\langle M \rangle \sim t^{\beta}, t > 0$.

We can follow Herrero and study the logarithmic derivative:
\begin{equation}
\mu(t) = \frac{d \log \langle M\rangle}{d \log t}
\end{equation}
This gives a qualitative indication of how the behaviour is changing as $p$ is varied.

The logarithmic derivative $\mu$ curves for different $p$ values appear to intercept the $t=0$ axis at values between a
3-D Ising-type transition $\beta$ value of $\approx 0.325$\cite{Gupta} and the mean-field value of $0.5$. Our 3-D data
has quite high uncertainties in it due to small numbers of samples with large system sizes, but it suggests a monotonic
increase in the limiting value of the $t=0$ intercept for $\mu$ with rising $p$.  Our 2-D data also shows a steady
change between the 2-D Ising value for $\beta$ of $0.125$ and the mean-field value.

\begin{figure}[hbt]
\begin{center}
\includegraphics[angle=-90,width=9cm]{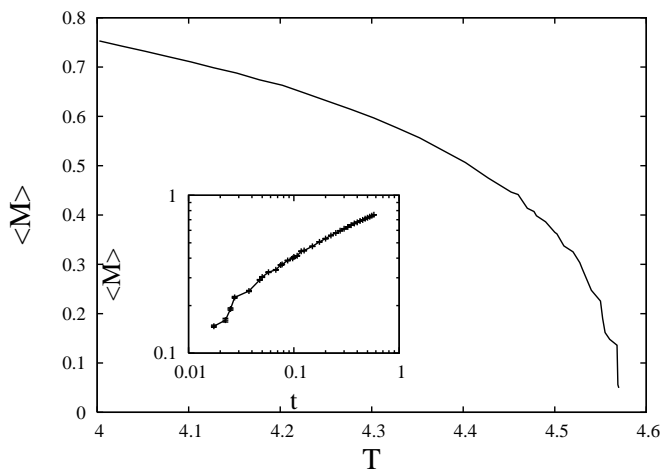}
\caption{\label{fig:mag} Deriving $\beta$ from the Magnetization }
\end{center}
\end{figure}

Figure~\ref{fig:mag} shows the magnetization averaged over $25$ sample sizes for a three-dimensional system at a finite
value of $p$ as we approach the critical temperature from below.  We can estimate the limiting value of the slope at low
$t$ (as shown in the inset) and use this to estimate values of $\beta$.

Applying this procedure we find a good straight line fit of  $\log \beta$ vs $\log p$ characterised by 
$\beta \approx p^{0.10 \pm 0.005}$  in 2-D and $\beta \approx p^{0.08 \pm 0.002}$ in 3-D.

We conclude that the model transitions continuously from Ising-like behaviour to mean-field behaviour as $p$ is
increased.  We find no difference in these values for the two different rewiring models.  This again emphasises the
power of the small-world parameter $p$ in interpolating between two different behaviour regimes.

\subsection*{Critical Exponent $\nu$}

The correlation length near the critical temperature  is known to scale as:
\begin{equation}
\xi \sim |T-T_c|^{-\nu}
\end{equation}
and the correlation length exponent $\nu$ is known to have values~\cite{KadanoffBook} of $\nu = 0.632$ for 3-D systems;
$\nu = 1$ for a 2-D system and a value of $\nu = 0.5$ from mean-field theory.

At the linearised region near the critical temperature, an expansion indicates that the Binder cumulant depends upon
approximately on temperature~\cite{KimOnXY} as:
\begin{equation}
U_N(T) \approx U^* + U_1 \left( 1-\frac{T}{T_C} \right) N^{\frac{1}{\nu}}
\end{equation}
so that 
\begin{equation}
\frac{\Delta U_N}{\Delta T} \propto -N^\frac{1}{\nu}
\end{equation}

On the work we report here on small-world  systems based upon regular lattices with a fixed $z$, we can also write:
\begin{equation}
\frac{\Delta U_N}{\Delta T} \propto -L^\frac{d}{\nu}
\end{equation}

Straight line fits to the experimental Binder cumulants around the critical temperature yields values of $\nu$ that are
between the requisite Ising value and mean-field values.  The experimental uncertainties in our data for $\nu$ are quite
high however, and although the values suggest a monotonic change between Ising and mean-field behaviour as $p$ is
increased, we are unable to identify a meaningful functional form to characterise the variation with $p$.

\subsection*{$p$-Dependence of the Transition}

It seems that the small-world transitional behaviour of the Ising model is intimately tied with enhancements to the
systems ability to support long-range correlations.  It is therefore useful to consider the length scales present in
the rewired lattice.  A useful metric is the maximum path length connecting two spin sites, as counted in terms of number of
traversed bonds or ``hops'' along edges of the graph.

\begin{figure}[hbt]
\begin{center}
\includegraphics[angle=-90,width=9cm]{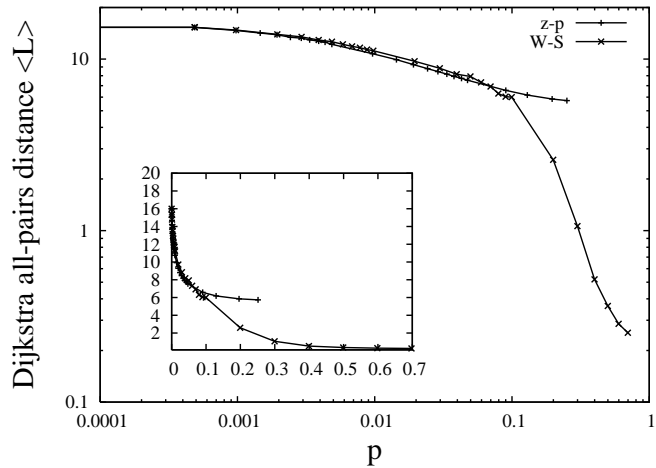}
\caption{\label{fig:dijkstra} Mean All-pairs Dijkstra distance $\langle L \rangle$ averaged over 25 samples of $32
  \times 32$ lattices perturbed with the Watts-Strogatz (W-S) and our $z$-preserving (z-p) rewiring models.}
\end{center}
\end{figure}

Figure~\ref{fig:dijkstra} shows the Mean All-pairs distance $\langle L \rangle$ calculated using the Dijkstra
algorithm~\cite{Dijkstra} for a 2-D lattice when perturbed with the Watts-Strogatz (W-S) and our $z$-preserving (z-p)
rewiring models.  The calculation is shown for a $32 \times 32$ system with the two different rewiring algorithms
applied.  The underpinning regular lattice is periodic so the furthest apart any two spin sites can be is $L / 2 =
16$ edge units.  This is shown on the inset plot on a linear scale, where the curves tend towards a maximum value of $16$
at $p=0$.

There are three distinct $p$-regimes shown in the figure.  At small values of $p$, for which the lattice size is big
enough to support a reasonable number of rewired bonds, the $\log-\log$ plot shows that $\langle L \rangle \approx
p^{m}$. We find $m \approx 0.20$ for this data set.  At $p$-values that are too small for the lattice to support, the
straight line tails off.  At high values of $p$ the value of $\langle L \rangle$ falls off to the value for a random
lattice.

Note that at high $p$ values, our $z$-preserving rewiring algorithm gives a different $\langle L \rangle$ value from the
W-S rewiring algorithm.  This was discussed in section~\ref{sec:rewiring} and is due to exclusion effects of our
algorithm in avoiding self-bonds and multiple bonds.  At high-$p$ the Watts-Strogatz rewiring algorithm breaks up the
system into multiple components.  This is reflected in the averaged Dijkstra distance calculations, and for the W-S
rewiring $\langle L \rangle$ falls to zero in the limit of a completely random lattice.

Herrero hypothesized  that  the order-disorder transition temperature for small $p$ becomes
\begin{equation}
T_c -  T_{c}^{p=0} \sim p^\frac{1}{\nu d}
\label{eq:hypothesis}
\end{equation}
where $\nu$ is the critical exponent of the regular (unperturbed) lattice.  His data supports this in a 2-D system, as
does ours.  However for a 3-D system his data disagree with this hypothesis and he speculated this was due to too small
system sizes and insufficiently small $p$.  This conclusion was also reached by \cite{KimOnXY} for the XY
model. Nevertheless our value of $s \approx 0.698$ for a 3-D system, for which we have quite a high degree of
confidence, and which is derived from much larger simulated system sizes, is still in disagreement
with equation~\ref{eq:hypothesis}.  Furthermore, in the case of our 4-D system, which has mean-field behaviour in the regular
unperturbed lattice case as well as in the rewired case, our measured value of $s = 0.75$ also clearly disagrees with
this hypothesis.

\section{Conclusions}
\label{sec:conclusions}

Generally the effect of the rewiring is to shift the the critical temperature $T_c$ upwards.  We have determined this
remains the case in dimensions $2, 3, 4$ and $5$.  We have also found that the precise local rewiring details are not
important to the nature of the small-world phase transition, providing the appropriate interpretation of parameter $p$
is made.  We find that the shift in $T_c$ seems to go as a power law in $p$ in all dimensions.

Since our re-wire model preserves $z$ and hence the effective dimensionality of the underpinning lattice, we believe it is this
effect that principally gives rise to the $p$-concentration dependences of $T_c$ and not the change in effective
dimensionality of the system that arises when links are added. The small-world shortcuts support long-range spin-spin
correlations above the normal $T_c$ value.  It appears that it is the presence of long-range correlations that
essentially constitute the nature of the phase transition and the critical behaviour; it is not unreasonable that the
small-worlding of the lattice has this effect.

Operationally we have found that a cluster-updating method such as that of Wolff is entirely satisfactory and its use
may assist in investigating more closely the small-world behaviour in large systems sizes at higher dimensionality.  The
exact nature of the small-world transition and in particular its interplay with the Ising transition are still not
entirely clear.  

We believe that the shift from Ising-like behaviour to mean-field behaviour is a gradual one, but it remains to
ascertain this with better statistical sampling in higher dimensional systems.  A particular area worthy of further
attention is the nature of cluster break-up in the system under a Watts-Strogatz rewiring, and the associated path and
correlation lengths that arise for different cluster components.

\begin{acknowledgments}
This work was funded by Massey University.  We thank P.D.Coddington for valuable discussions on use of the
Wolff cluster method.
\end{acknowledgments}

\bibliography{small-world}

\begin{thebibliography}{28}
\expandafter\ifx\csname natexlab\endcsname\relax\def\natexlab#1{#1}\fi
\expandafter\ifx\csname bibnamefont\endcsname\relax
  \def\bibnamefont#1{#1}\fi
\expandafter\ifx\csname bibfnamefont\endcsname\relax
  \def\bibfnamefont#1{#1}\fi
\expandafter\ifx\csname citenamefont\endcsname\relax
  \def\citenamefont#1{#1}\fi
\expandafter\ifx\csname url\endcsname\relax
  \def\url#1{\texttt{#1}}\fi
\expandafter\ifx\csname urlprefix\endcsname\relax\def\urlprefix{URL }\fi
\providecommand{\bibinfo}[2]{#2}
\providecommand{\eprint}[2][]{\url{#2}}

\bibitem[{\citenamefont{Watts and Strogatz}(1998)}]{Watts+Strogatz}
\bibinfo{author}{\bibfnamefont{D.~J.} \bibnamefont{Watts}} \bibnamefont{and}
  \bibinfo{author}{\bibfnamefont{S.~H.} \bibnamefont{Strogatz}},
  \bibinfo{journal}{Nature} \textbf{\bibinfo{volume}{393}},
  \bibinfo{pages}{440} (\bibinfo{year}{1998}).

\bibitem[{\citenamefont{Klemm et~al.}(2003)\citenamefont{Klemm, Eguiluz, Toral,
  and Miguel}}]{Klemm-SocialNetworks}
\bibinfo{author}{\bibfnamefont{K.}~\bibnamefont{Klemm}},
  \bibinfo{author}{\bibfnamefont{V.~M.} \bibnamefont{Eguiluz}},
  \bibinfo{author}{\bibfnamefont{R.}~\bibnamefont{Toral}}, \bibnamefont{and}
  \bibinfo{author}{\bibfnamefont{M.~S.} \bibnamefont{Miguel}},
  \bibinfo{journal}{Phys.\ Rev. \ E} \textbf{\bibinfo{volume}{67}},
  \bibinfo{pages}{026120} (\bibinfo{year}{2003}).

\bibitem[{\citenamefont{Wan}(2005)}]{WanOnSmallWorldLQG}
\bibinfo{author}{\bibfnamefont{Y.}~\bibnamefont{Wan}} (\bibinfo{year}{2005}),
  \bibinfo{note}{arXiv.org:hep-th/0512210}.

\bibitem[{\citenamefont{Newman}(2003)}]{NewmanComplexNetworks}
\bibinfo{author}{\bibfnamefont{M.~E.~J.} \bibnamefont{Newman}},
  \bibinfo{journal}{SIAM Review} \textbf{\bibinfo{volume}{45}},
  \bibinfo{pages}{169} (\bibinfo{year}{2003}).

\bibitem[{\citenamefont{Bollobas}(1985)}]{Bollobas}
\bibinfo{author}{\bibfnamefont{B.}~\bibnamefont{Bollobas}},
  \emph{\bibinfo{title}{Random Graphs}} (\bibinfo{publisher}{Academic Press,
  New York}, \bibinfo{year}{1985}).

\bibitem[{\citenamefont{Mathias and Gopal}(2001)}]{Mathias+Gopal}
\bibinfo{author}{\bibfnamefont{N.}~\bibnamefont{Mathias}} \bibnamefont{and}
  \bibinfo{author}{\bibfnamefont{V.}~\bibnamefont{Gopal}},
  \bibinfo{journal}{Phys. \ Rev. \ E} p. \bibinfo{pages}{021117}
  (\bibinfo{year}{2001}).

\bibitem[{\citenamefont{Newman}(2000)}]{NewmanSmallWorldModels}
\bibinfo{author}{\bibfnamefont{M.~E.~J.} \bibnamefont{Newman}}
  (\bibinfo{year}{2000}), \bibinfo{note}{arXiv.org:cond-mat/0001118}.

\bibitem[{\citenamefont{Loes et~al.}(2004)\citenamefont{Loes, Pogorelov, dos
  Santos, and Toral}}]{LopesExactIsing}
\bibinfo{author}{\bibfnamefont{J.~V.} \bibnamefont{Loes}},
  \bibinfo{author}{\bibfnamefont{Y.~G.} \bibnamefont{Pogorelov}},
  \bibinfo{author}{\bibfnamefont{J.~M. B.~L.} \bibnamefont{dos Santos}},
  \bibnamefont{and} \bibinfo{author}{\bibfnamefont{R.}~\bibnamefont{Toral}},
  \bibinfo{journal}{Phys.\ Rev. \ E} \textbf{\bibinfo{volume}{70}},
  \bibinfo{pages}{026112} (\bibinfo{year}{2004}).

\bibitem[{\citenamefont{Herrero}(2002)}]{Herrero}
\bibinfo{author}{\bibfnamefont{C.~P.} \bibnamefont{Herrero}},
  \bibinfo{journal}{Phys.\ Rev.\ E} \textbf{\bibinfo{volume}{65}},
  \bibinfo{pages}{066110} (\bibinfo{year}{2002}).

\bibitem[{\citenamefont{Herrero}(2004)}]{Herrero2}
\bibinfo{author}{\bibfnamefont{C.~P.} \bibnamefont{Herrero}},
  \bibinfo{journal}{Phys.\ Rev.\ E} \textbf{\bibinfo{volume}{69}},
  \bibinfo{pages}{067109} (\bibinfo{year}{2004}).

\bibitem[{\citenamefont{Hastings}(2003)}]{HastingsSmallWorld}
\bibinfo{author}{\bibfnamefont{M.~B.} \bibnamefont{Hastings}},
  \bibinfo{journal}{Phys.\ Rev.\ Lett.} \textbf{\bibinfo{volume}{91}},
  \bibinfo{pages}{098701} (\bibinfo{year}{2003}).

\bibitem[{\citenamefont{Zhang and Novotny}(2006)}]{Zhang+NovotnyOnLongRange}
\bibinfo{author}{\bibfnamefont{X.}~\bibnamefont{Zhang}} \bibnamefont{and}
  \bibinfo{author}{\bibfnamefont{M.~A.} \bibnamefont{Novotny}}
  (\bibinfo{year}{2006}), \bibinfo{note}{arXiv.org:cond-mat/0602286}.

\bibitem[{\citenamefont{Novotny et~al.}(2004)\citenamefont{Novotny, Zhang,
  Yancey, Dubreus, Cook, Gill, Norwood, and
  Novotny}}]{NovotnyEtAlOnNanomaterials}
\bibinfo{author}{\bibfnamefont{A.~M.} \bibnamefont{Novotny}},
  \bibinfo{author}{\bibfnamefont{X.}~\bibnamefont{Zhang}},
  \bibinfo{author}{\bibfnamefont{J.}~\bibnamefont{Yancey}},
  \bibinfo{author}{\bibfnamefont{T.}~\bibnamefont{Dubreus}},
  \bibinfo{author}{\bibfnamefont{M.~L.} \bibnamefont{Cook}},
  \bibinfo{author}{\bibfnamefont{S.~G.} \bibnamefont{Gill}},
  \bibinfo{author}{\bibfnamefont{I.~T.} \bibnamefont{Norwood}},
  \bibnamefont{and} \bibinfo{author}{\bibfnamefont{A.~M.}
  \bibnamefont{Novotny}} (\bibinfo{year}{2004}),
  \bibinfo{note}{arXiv.org:cond-mat/0410589}.

\bibitem[{\citenamefont{Onsager}(1944)}]{Onsager}
\bibinfo{author}{\bibfnamefont{L.}~\bibnamefont{Onsager}},
  \bibinfo{journal}{Phys.Rev.} \textbf{\bibinfo{volume}{65}},
  \bibinfo{pages}{117} (\bibinfo{year}{1944}).

\bibitem[{\citenamefont{Baillie et~al.}(1992)\citenamefont{Baillie, Gupta,
  Hawick, and Pawley}}]{MCRG}
\bibinfo{author}{\bibfnamefont{C.~F.} \bibnamefont{Baillie}},
  \bibinfo{author}{\bibfnamefont{R.}~\bibnamefont{Gupta}},
  \bibinfo{author}{\bibfnamefont{K.~A.} \bibnamefont{Hawick}},
  \bibnamefont{and} \bibinfo{author}{\bibfnamefont{G.~S.}
  \bibnamefont{Pawley}}, \bibinfo{journal}{Phys.\ Rev. \ B}
  \textbf{\bibinfo{volume}{45}}, \bibinfo{pages}{10438} (\bibinfo{year}{1992}).

\bibitem[{\citenamefont{Svenson and Johnston}(2002)}]{Svenson+Johnson}
\bibinfo{author}{\bibfnamefont{P.}~\bibnamefont{Svenson}} \bibnamefont{and}
  \bibinfo{author}{\bibfnamefont{D.~A.} \bibnamefont{Johnston}},
  \bibinfo{journal}{Phys.\ Rev. \ E} \textbf{\bibinfo{volume}{65}},
  \bibinfo{pages}{036105} (\bibinfo{year}{2002}).

\bibitem[{\citenamefont{Wolff}(1989)}]{Wolff}
\bibinfo{author}{\bibfnamefont{U.}~\bibnamefont{Wolff}},
  \bibinfo{journal}{Phys.\ Lett.} \textbf{\bibinfo{volume}{228}},
  \bibinfo{pages}{379} (\bibinfo{year}{1989}).

\bibitem[{\citenamefont{Binney et~al.}(1992)\citenamefont{Binney, Dowrick,
  Fisher, and Newman}}]{CriticalPhenomena}
\bibinfo{author}{\bibfnamefont{J.~J.} \bibnamefont{Binney}},
  \bibinfo{author}{\bibfnamefont{N.~J.} \bibnamefont{Dowrick}},
  \bibinfo{author}{\bibfnamefont{A.~J.} \bibnamefont{Fisher}},
  \bibnamefont{and} \bibinfo{author}{\bibfnamefont{M.~E.~J.}
  \bibnamefont{Newman}}, \emph{\bibinfo{title}{The Theory of Critical
  Phenomena}} (\bibinfo{publisher}{Oxford University Press},
  \bibinfo{year}{1992}).

\bibitem[{\citenamefont{Meilikhov and
  Farzetdinova}(2005)}]{Meilikhov+Farzetdinova}
\bibinfo{author}{\bibfnamefont{E.~Z.} \bibnamefont{Meilikhov}}
  \bibnamefont{and} \bibinfo{author}{\bibfnamefont{R.~M.}
  \bibnamefont{Farzetdinova}} (\bibinfo{year}{2005}),
  \bibinfo{note}{arxiv.org/abs/cond-mat/0505502}.

\bibitem[{\citenamefont{Glauber}(1963)}]{Glauber}
\bibinfo{author}{\bibfnamefont{R.}~\bibnamefont{Glauber}}, \bibinfo{journal}{J.
  Math. Phys II} pp. \bibinfo{pages}{294--307} (\bibinfo{year}{1963}).

\bibitem[{\citenamefont{Marsaglia and Zaman}(1987)}]{Marsaglia}
\bibinfo{author}{\bibfnamefont{G.}~\bibnamefont{Marsaglia}} \bibnamefont{and}
  \bibinfo{author}{\bibfnamefont{A.}~\bibnamefont{Zaman}}
  (\bibinfo{year}{1987}), \bibinfo{note}{{FSU-SCRI-87-50}, Florida State
  University}.

\bibitem[{\citenamefont{Matsumoto and Nishimura}(1998)}]{MersenneTwistor}
\bibinfo{author}{\bibfnamefont{M.}~\bibnamefont{Matsumoto}} \bibnamefont{and}
  \bibinfo{author}{\bibfnamefont{T.}~\bibnamefont{Nishimura}},
  \bibinfo{journal}{ACM Transactions on Modeling and Computer Simulation}
  \textbf{\bibinfo{volume}{8 No 1.}}, \bibinfo{pages}{3}
  (\bibinfo{year}{1998}).

\bibitem[{\citenamefont{Binder and Landau}(1984)}]{Binder+LandauCumulant}
\bibinfo{author}{\bibfnamefont{K.}~\bibnamefont{Binder}} \bibnamefont{and}
  \bibinfo{author}{\bibfnamefont{D.~P.} \bibnamefont{Landau}},
  \bibinfo{journal}{Phys.\ Rev.\ B} \textbf{\bibinfo{volume}{30/3}},
  \bibinfo{pages}{1477} (\bibinfo{year}{1984}).

\bibitem[{\citenamefont{Binder and Heermann}(1997)}]{Binder+Heermann}
\bibinfo{author}{\bibfnamefont{K.}~\bibnamefont{Binder}} \bibnamefont{and}
  \bibinfo{author}{\bibfnamefont{D.~W.} \bibnamefont{Heermann}},
  \emph{\bibinfo{title}{Monte Carlo Simulation in Statistical Physics}}
  (\bibinfo{publisher}{Springer-Verlag}, \bibinfo{year}{1997}).

\bibitem[{\citenamefont{Gupta and Tamayo}(1996)}]{Gupta}
\bibinfo{author}{\bibfnamefont{R.}~\bibnamefont{Gupta}} \bibnamefont{and}
  \bibinfo{author}{\bibfnamefont{P.}~\bibnamefont{Tamayo}}
  (\bibinfo{year}{1996}), \bibinfo{note}{arXiv:cond-mat/9601048}.

\bibitem[{\citenamefont{Kadanoff}(2000)}]{KadanoffBook}
\bibinfo{author}{\bibfnamefont{L.~P.} \bibnamefont{Kadanoff}},
  \emph{\bibinfo{title}{Statistical Physics Statics, Dynamics and
  Renormalization}}, \bibinfo{number}{981-02-3764} (\bibinfo{publisher}{World
  Scientific}, \bibinfo{year}{2000}).

\bibitem[{\citenamefont{Kim et~al.}(2001)\citenamefont{Kim, Hong, Holme, Jeon,
  Minnhagen, and Choi}}]{KimOnXY}
\bibinfo{author}{\bibfnamefont{B.~J.} \bibnamefont{Kim}},
  \bibinfo{author}{\bibfnamefont{H.}~\bibnamefont{Hong}},
  \bibinfo{author}{\bibfnamefont{P.}~\bibnamefont{Holme}},
  \bibinfo{author}{\bibfnamefont{G.~S.} \bibnamefont{Jeon}},
  \bibinfo{author}{\bibfnamefont{P.}~\bibnamefont{Minnhagen}},
  \bibnamefont{and} \bibinfo{author}{\bibfnamefont{M.~Y.} \bibnamefont{Choi}},
  \bibinfo{journal}{Phys.\ Rev. \ E} \textbf{\bibinfo{volume}{64}},
  \bibinfo{pages}{056135} (\bibinfo{year}{2001}).

\bibitem[{\citenamefont{Dijkstra}(1959)}]{Dijkstra}
\bibinfo{author}{\bibfnamefont{E.~W.} \bibnamefont{Dijkstra}},
  \bibinfo{journal}{Numerische Mathematik} \textbf{\bibinfo{volume}{1}},
  \bibinfo{pages}{269} (\bibinfo{year}{1959}).

\end{thebibliography}
\end{document}